# LASSO: A New Tool to Support Instructors and Researchers

*Ben Van Dusen, California State University Chico*

We developed the Learning About STEM Student Outcomes (LASSO) online assessment platform to increase instructor use of research-based assessments (RBAs). LASSO does this by making it easy to collect and analyze high-quality evidence about student learning in their courses. Specifically, LASSO simplifies the process of administering, scoring, and analyzing RBAs and saves class time by automating the process online. Course results are anonymized and aggregated in the LASSO database to provide instructors normative feedback about their student outcomes.

RBAs, such as the Force Concept Inventory, measure students' knowledge of concepts or attitudes that are core to a discipline. The LASSO database offers researchers access to a large-scale, multi-disciplinary, and longitudinal student and course-level data. The database can save researchers significant time and allow them to investigate novel research questions that require large datasets.

In this article we will discuss: 1) how LASSO supports instructors, 2) how LASSO supports researchers, and 3) research on collecting and analyzing data using LASSO.

## LASSO Supports Instructors

To measure student changes in STEM courses, the LASSO platform hosts, administers, scores, and analyzes student pretest and posttest scores online. Figure 1 outlines the steps for instructors to use LASSO. The LASSO platform is hosted on the Learning Assistant (LA) Alliance website.[1]

Instructors add new courses by answering a short series of questions about their course. Instructors then select assessments from the LASSO repository to administer to their students. As of the Fall '18 term, LASSO hosts sixteen research-based conceptual and attitudinal assessments across the STEM disciplines. Once instructors upload a course roster with emails and select a deadline for the pretest, they can launch the pretest. Each student receives an email with participation instructions including a personalized link to their online assessment. Students first choose whether they would like their answers to be anonymized and aggregated into the LASSO research database. They then complete a set of demographics questions and the RBA.

After students have completed their pretests, instructors can download a spreadsheet of their students' raw and scored responses. They can use the student responses to inform teaching practices, such as identifying concepts the students are more knowledgeable about, identifying students who may need additional support, and creating student small groups.

During the final weeks of the course, instructors follow the same steps for launching and tracking their students' progress on the posttests as they did on the pretest. Instructors can then download a spreadsheet with their students' pre and posttest responses as well as a final report. The spreadsheet supports faculty who wish to research their own course outcomes or upload their results to another data analysis system, such as Data Explorer. The final report is an assessment-specific PDF that provides instructors with an easy-to-understand analysis about their class's performance.

## LASSO supports research

The LASSO Platform aggregates and anonymizes the assessment data for researchers with IRB approval to use. Most students who take part in LASSO assessments (83%) agree to share their anonymized data with researchers. Besides providing researchers with information about student performance and demographics, the database also provides course-level information (e.g., goals of the course, how many times the instructor has taught the course before, and the class size). As of the Summer 2018 term, the LASSO research database has data from 32,728 students, in 618 courses, from 51 institutions (Table 1).

| Discipline | Institutions | Instructors | Courses | Students |
|---|---|---|---|---|
| Physics | 41 | 129 | 462 | 19,819 |
| Astronomy | 3 | 3 | 3 | 181 |
| Mathematics | 7 | 11 | 30 | 2,257 |
| Chemistry | 12 | 20 | 68 | 5,764 |
| Biology | 12 | 21 | 75 | 5,575 |

*Table 1.* Data within the LASSO researcher database by discipline as of the 2018 Fall term.

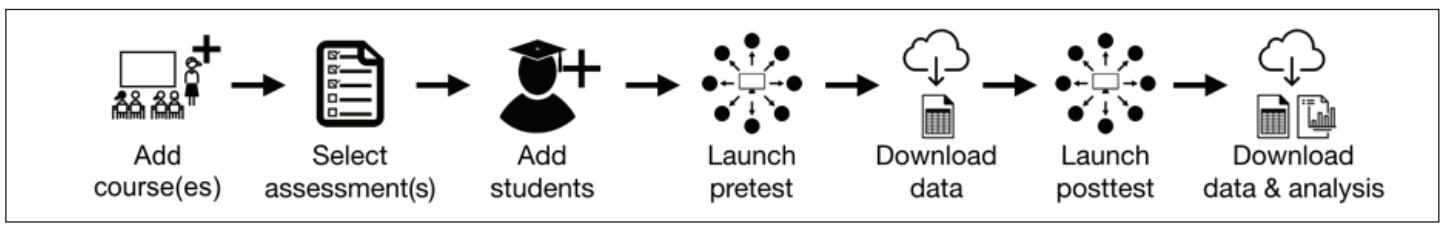

*Figure 1*. Steps to assessing a course using the LASSO platform.

While all instructor features on LASSO are free, there are fees to access to the LASSO database. The fees are small enough to not prevent researcher access to the database while providing funds to make the LASSO platform sustainable.

### Research on LASSO

We developed LASSO to support educators and researchers in collecting high quality data using instruments and analyses with strong validity arguments. To support this goal, we investigated two research questions of interest to LASSO-using instructors and researchers:

1. Are online assessments a good replacement for paper assessments?
2. What are the best methods for handling missing data?

We also investigated a third research question specifically for researchers

3. What are the best methods for analyzing large-scale multilevel datasets?

### Are online assessments a good replacement for paper assessments?

Nissen et al.[2] used a randomized between groups experimental design to investigate whether LASSO administered RBAs provided equivalent data to traditional in-class assessments for both student performance and participation. Analysis of 1,310 students in 3 college physics courses indicated that LASSO-based and in-class assessments provide equivalent participation rates when instructors used four recommended practices (shown in figure 2): (1) In-class reminders, (2) multiple email reminders, and (3) credit for pretest participation, and (4) credit for posttest participation. Models of student performance indicated that tests administered with LASSO had equivalent scores to those administered in class. This indicates that instructors can compare their data from LASSO to any prior data they may have collected and the broader literature on student gains.

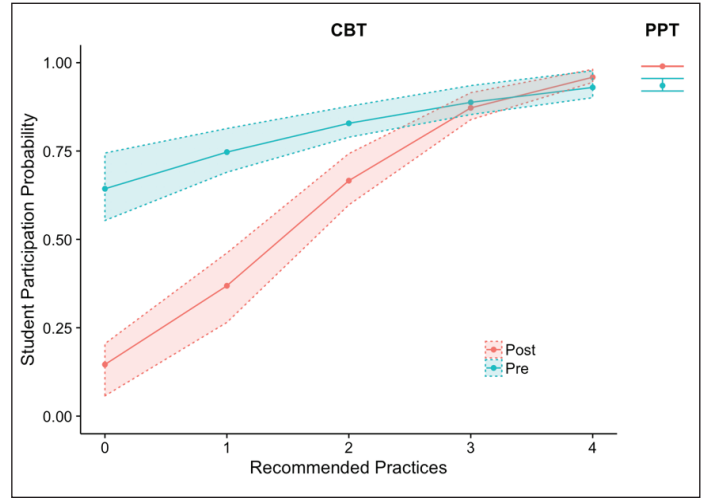

*Figure 2.* Participation rates on LASSO as instructors increased their use of the recommended practices (e.g., sending email reminders & offering credit) on computer-based tests (CBT) versus paper and pencil tests (PPT). When all 4 recommended practices were used, the participation rates were nearly identical.

### What are the best methods for handling missing data?

Nissen et al.[2] found that students with lower grades participated at lower rates than students with higher grades. These results indicated a bias toward high performing students for RBAs collected in-class or with LASSO. PER studies most commonly report using complete-case analysis (aka, matched data) in which data is discarded for any student who does not complete both the pre and posttest. Nissen, Donatello, and Van Dusen[3] used simulated classroom data to measure the potential bias introduced by complete case analysis and Multiple Imputation. Multiple Imputation uses all of the available data to build statistical models, which allows it to account for patterns in the missing data. Results, shown in Figure 3, indicated that complete-case analysis introduced meaningfully more bias into the results than multiple imputation.

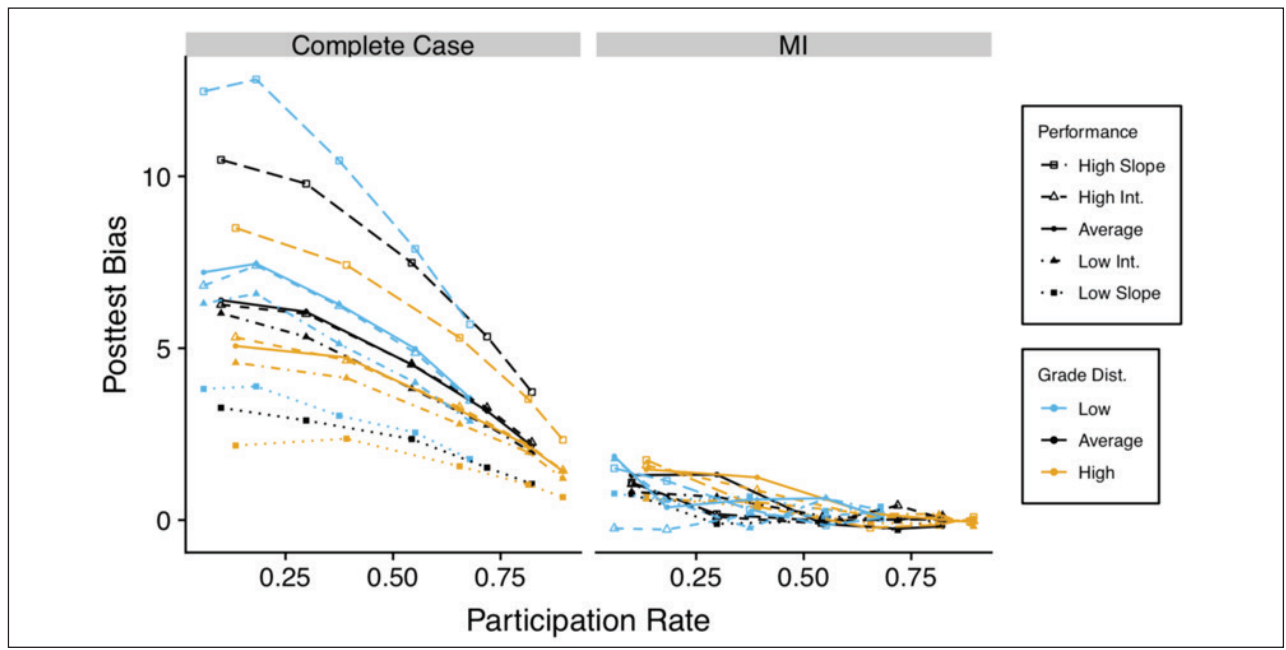

*Figure 3.* Bias introduced into posttest scores for complete case analysis and multiple imputation.

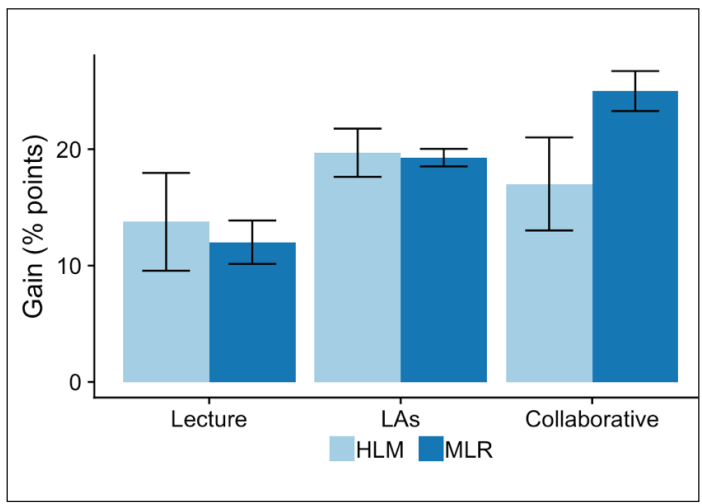

*Figure 4. Predicted gains for average students across course contexts as predicted by: a) multiple linear regression and b) hierarchical linear modeling. Error bars are +/- 1 standard error.*

**What are the best methods for analyzing large-scale multi-level databases?**

PER studies often use single-level regression models (e.g., linear and logistic regression) to analyze student outcomes. However, education datasets often have hierarchical structures, such as students nested within courses, that single-level models fail to account for. Multi-level models account for the structure of hierarchical datasets.

To illustrate the importance of performing a multi-level analysis of nested data, Van Dusen and Nissen[3] analyzed a dataset with 112 introductory physics courses from the LASSO database using both multiple linear regression and hierarchical linear modeling. They developed models that examined student learning in classrooms that use traditional instruction, collaborative learning with LAs, and collaborative learning without LAs. The two models produced significantly different findings about the impact of courses that used collaborative learning without LAs, shown in Figure 4. This analysis illustrated that the use of multi-level models to analyze nested datasets can impact the findings and implications of studies in PER. They concluded that the DBER community should use multi-level models to analyze datasets with hierarchical structures.

**Conclusion**

The LASSO platforms purpose is to support instructors in implementing research-based teaching practices in their courses by providing them with simple, accurate, and reliable assessments for their courses and to support research on STEM instruction. The LASSO platform makes it easy for instructors to assess their courses, supports instructors interpreting the results from their assessments, and provides them with documentation summarizing their assessment results. Large-scale, multi-disciplinary data collection allows researchers to further understanding of student learning in STEM.

*Ben Van Dusen is an assistant professor in Science Education at Chico State and the director of the LASSO platform.*

**(Endnotes)**

1. www.learningassistantalliance.org
2. J. M. Nissen, M. Jariwala, E. W. Close, & B. Van Dusen, "Participation and performance on paper-and computer-based low-stakes assessments," *International Journal of STEM Education*, **5**(1), 21, (2018).
3. J. Nissen, R. Donatello, & B. Van Dusen, "Missing data and bias in physics education research: A case for using multiple imputation," *Physical Review Physics Education Research* (under review).
4. Van Dusen and Nissen "Modernizing PER's use of regression models: a review of hierarchical linear modeling," *Physical Review Physics Education Research* (under review).